\documentclass{aa}  

\usepackage{graphicx}
\usepackage{txfonts}
\begin{document}

	\title{Gaia18aen: First symbiotic star discovered by \textit{Gaia}}
	
	
	\author{J.~Merc
		\inst{1,2}\thanks{\email{jaroslav.merc@student.upjs.sk}}
		\and
		J.~Miko\l{}ajewska\inst{3}
		\and
		M.~Gromadzki\inst{4}
		\and
		C.~Ga\l{}an\inst{3}
		\and
		K.~I\l{}kiewicz\inst{3,5}
		\and
		J.~Skowron\inst{4}
		\and
		{\L}.~Wyrzykowski\inst{4}
		\and 
		S.~T.~Hodgkin\inst{6}
		\and
		K.~A.~Rybicki\inst{4}
		\and
		P.~Zieliński\inst{4}
		\and
		K.~Kruszyńska\inst{4}
		\and
		V.~Godunova \inst{7}
		\and
		A.~Simon \inst{8}
		\and
		V.~Reshetnyk \inst{8}
		\and
		F.~Lewis \inst{9,10}
		\and
		U.~Kolb \inst{11}
		\and
		M.~Morrell \inst{11}
		\and
		A.~J.~Norton \inst{11}
		\and
		S.~Awiphan \inst{12}
		\and
		S.~Poshyachinda \inst{12}
		\and
		D.~E.~Reichart \inst{13}
		\and
		M.~Greet \inst{14}
		\and
		J.~Kolgjini \inst{14}
	}
	
	\institute{Astronomical Institute, Faculty of Mathematics and Physics, Charles University, V Hole\v{s}ovi\v{c}k{\'a}ch 2, 180 00 Prague, Czechia
		\and
		Institute of Physics, Faculty of Science, P. J. \v{S}af{\'a}rik University, Park Angelinum 9, 040 01 Ko\v{s}ice, Slovakia
		\and
		Nicolaus Copernicus Astronomical Center, Polish Academy of Sciences, Bartycka 18, 00–716 Warsaw, Poland
		\and
		Astronomical Observatory, University of Warsaw, Al. Ujazdowskie 4, 00-478 Warszaw, Poland
		\and
		Department of Physics and Astronomy, Box 41051, Science Building, Texas Tech University, Lubbock, TX 79409-1051, USA
		\and
		Institute of Astronomy, University of Cambridge, Madingley Road CB3 0HA, Cambridge, UK
		\and
		ICAMER Observatory of NASU, 27 Acad. Zabolotnoho str., Kyiv, 03143, Ukraine
		\and
		Faculty of Physics, Taras Shevchenko National University of Kyiv, 4 Glushkova Ave., Kyiv, 03022, Ukraine
		\and
		Faulkes Telescope Project, School of Physics, and Astronomy, Cardiff University, The Parade, Cardiff CF24 3AA, UK
		\and
		Astrophysics Research Institute, Liverpool John Moores University, 146 Brownlow Hill, Liverpool L3 5RF, UK
		\and
		School of Physical Sciences, The Open University, Walton Hall, Milton Keynes MK7 6AA, UK
		\and
		National Astronomical Research Institute of Thailand, 260, Moo 4, T. Donkaew, A. Mae Rim, Chiang Mai, 50180, Thailand
		\and
		Department of Physics and Astronomy, University of North Carolina at Chapel Hill, Chapel Hill, NC 27599, USA
		\and
		Eastbury Community School, Hulse Avenue, Barking IG11 9UW, UK
	}
	
	\date{Received \today; accepted \today}
	
	
	\abstract
	{Besides the astrometric mission of the \textit{Gaia} satellite, its repeated and high-precision measurements serve also as an all-sky photometric transient survey. The sudden brightenings of the sources are published as Gaia Photometric Science Alerts and are made publicly available allowing the community to photometrically and spectroscopically follow-up the object.}
	{The goal of this paper was to analyze the nature and derive the basic parameters of Gaia18aen, transient detected at the beginning of 2018. It coincides with the position of the emission line star WRAY 15-136. The brightening was classified as a "nova?" on the basis of subsequent spectroscopic observation.}
	{We have analyzed two spectra of Gaia18aen and collected the available photometry of the object covering the brightenings in 2018 and also the preceding and following periods of quiescence. Based on this observational data, we have derived the parameters of Gaia18aen and discussed the nature of the object.}
	{Gaia18aen is the first symbiotic star discovered by \textit{Gaia} satellite. The system is an S-type symbiotic star and consists of an M giant of a slightly super-solar metallicity, with T$_{\rm eff}$\,$\sim$\,3\,500\,K, a radius of $\sim$\,230\,R$\sun$, and a high luminosity L\,$\sim$\,7\,400\,L$\sun$. The hot component is a hot white dwarf. We tentatively determined the orbital period of the system $\sim$\,487 days. The main outburst of Gaia18aen in 2018 was accompanied by a decrease in the temperature of the hot component. The first phase of the outburst was characterized by the high luminosity L\,$\sim$\,27\,000\,L$\sun$, which remained constant for about three weeks after the optical maximum, later followed by the gradual decline of luminosity and increase of temperature. Several re-brightenings have been detected on the timescales of hundreds of days.}
	{}
	
	
	\maketitle
	%
	
	\section{Introduction}
	Symbiotic stars are among the widest interacting binaries. They are consisting of a cool giant (or a supergiant) of a spectral type M (in yellow symbiotics K, rarely G) as the donor and a compact star, the most commonly a hot white dwarf ($\sim10^5$\,K), as the accretor \citep{2007BaltA..16....1M}. These binaries are usually embedded in the circumbinary nebula created by the winds of both components. Thanks to their properties, symbiotic stars can serve as unique astrophysical laboratories to study accretion processes, winds, or jets \citep[for more information see e.g. reviews by][]{2012BaltA..21....5M, 2019arXiv190901389M}.
	
	Most of the symbiotic stars in the previous century were discovered serendipitously as a result of their strong outbursts or during the spectroscopic surveys based on their peculiar spectral appearance. In the recent years, several surveys focused especially on looking for new symbiotic stars - in the Milky Way \citep[e.g.][]{2013MNRAS.432.3186M, 2014MNRAS.440.1410M,2014A&A...567A..49R} and in the external galaxies \citep[e.g.][]{2008MNRAS.391L..84G, 2012MNRAS.419..854G, 2015MNRAS.447..993G, 2009MNRAS.395.1121K, 2014MNRAS.444..586M,2017MNRAS.465.1699M,2018arXiv181106696I}.
	
	Here we report on the first discovery of the symbiotic star, Gaia18aen, by the \textit{Gaia} satellite. \object{Gaia18aen} (AT 2018id, WRAY 15-136) was previously classified as an emission line star by \citet{1966PhDT.........3W}. Its outburst was detected by the \textit{Gaia} satellite and announced by the Gaia Science Alert\footnote{http://gsaweb.ast.cam.ac.uk/alerts/home} \citep[GSA;][]{2012IAUS..285..425W,2013RSPTA.37120239H,2014gfss.conf...31W} on January 17, 2018 \citep{2018TNSTR..84....1D}, when the star had the magnitude $G$ = 11.33. It was referred to as a "bright emission-line star in Galactic plane which brightened by 1 magnitude" in the alert. Previous measurements of the \textit{Gaia} satellite over the period from October 31, 2014, to November 3, 2017, shown the average magnitude of the star was 12.31 $\pm$ 0.10 with no significant changes. According to \textit{Gaia} data, the star started to increase its brightness at the turn of November and December 2017. Observation obtained on December 3, 2017, revealed the star at the magnitude of 12.07 and object continued to brighten in the following weeks. \citet{2018ATel11634....1K} suggested a "nova?" classification for the object based on the spectrum obtained by VLT/X-Shooter as a part of the program focused on spectroscopic classification of candidates for microlensing events. As it is discussed in this paper, the observed event was not a nova outburst, but a Z And-type outburst of a classical symbiotic star. 
	
	This paper is organized as followed: in Section \ref{sec:observations} we discuss observational data which were used for the classification of the object and analysis of its behavior and parameters, in Section \ref{sec:results} we describe spectra and light curves of the object, parameters of the symbiotic components and discuss its variability and outburst activity. 
	
	\section{Observations}\label{sec:observations}
	
	\begin{table}
		\caption{Basic properties of Gaia18aen. Data are from Gaia DR2 \citep{2018A&A...616A...1G}, 2MASS \citep{2006AJ....131.1163S}, and WISE \citep{2010AJ....140.1868W}.}           
		\label{tab:gaia}      
		\centering                          
		\begin{tabular}{c c}        
			\hline\hline    
			Parameter               & Value                   \\\hline
			$\alpha_{2000}$ [h:m:s] & 08:02:52.06             \\
			$\delta_{2000}$ [d:m:s] & -30:18:37.19            \\
			l$_{2000}$ [deg]          & 247.674                 \\
			b$_{2000}$ [deg]           & +0.314                  \\
			$\mu_\alpha$ [mas/yr]    & -1.691 $\pm$ 0.084        \\
			$\mu_\delta$ [mas/yr]    & 4.140 $\pm$ 0.078         \\
			$\pi$ [mas]          & 0.097 $\pm$ 0.054         \\
			$G$ [mag]                 & 12.441 $\pm$ 0.003        \\
			$BP$ [mag]                & 14.481 $\pm$ 0.019        \\
			$RP$ [mag]                 & 11.089 $\pm$ 0.006 \\
			$J$ [mag] & 8.515 $\pm$ 0.027 \\
			$H$ [mag] & 7.189 $\pm$ 0.047 \\
			$K\rm _S$ [mag] & 6.680 $\pm$ 0.036 \\
			$W1$ [mag] & 6.369 $\pm$ 0.042 \\
			$W2$ [mag] & 6.359 $\pm$ 0.022 \\
			$W3$ [mag] & 6.112 $\pm$ 0.014 \\
			$W4$ [mag] & 5.817 $\pm$ 0.038 
		\end{tabular}
	\end{table}
	
	\subsection{Spectroscopy}
	
	We collected two spectroscopic observations of Gaia18aen. The first is a sequence of three low-resolution spectra obtained with the SPectrograph for the Rapid Acquisition of Transients (SPRAT) mounted on the Liverpool Telescope at La Palma \citep{2004SPIE.5489..679S} on January 20, 2018, under the program XOL17B02 (PI: Hodgkin). The exposure time of each spectrum is 30 s. They have a wavelength range of 4\,000 -- 8\,000 \AA\ and the resolution $R \sim 350$. Spectra were extracted, wavelength calibrated, and flux calibrated using the SPRAT pipeline \citep{2014SPIE.9147E..8HP}. In our analysis we used averaged spectrum. 
	
	Another spectroscopic observation was obtained by VLT/X-Shooter \citep{2011A&A...536A.105V}. Two exposures were obtained in the A-B nodding mode on March 22, 2018, under the program 0100.D-0021 (PI: Wyrzykowski). We used slits width 1.0”, 0.7”, 0.6” what produced resolutions of 4\,300, 11\,000, 7\,900, single exposures times were 91, 120, 10s in the UVB (2\,989 - 5\,560 \AA), VIS (5\,337 - 10\,200 \AA), and NIR (9\,940 - 24\,790 \AA) arms, respectively. We reduced the spectra with the dedicated EsoReflex pipeline (v. 3.3.4). The spectrum was corrected for telluric features using the {\sl{MolecFit}} package \citep{2015A&A...576A..78K,2015A&A...576A..77S}.  We found this tool the most efficient to work with VLT/X-Shooter data \citep[see][]{2019A&A...621A..79U}. The standard procedure was applied similar to those described for the VLT/X-Shooter observations by \citet{2015A&A...576A..78K} through fitting to the atmospheric absorption features from molecules of H$_2$O and O$_2$ in the visual range, and additionally CO$_2$ and CH$_4$ in near-infrared.
	
\begin{figure}
		\centering
		\includegraphics[width=\columnwidth]{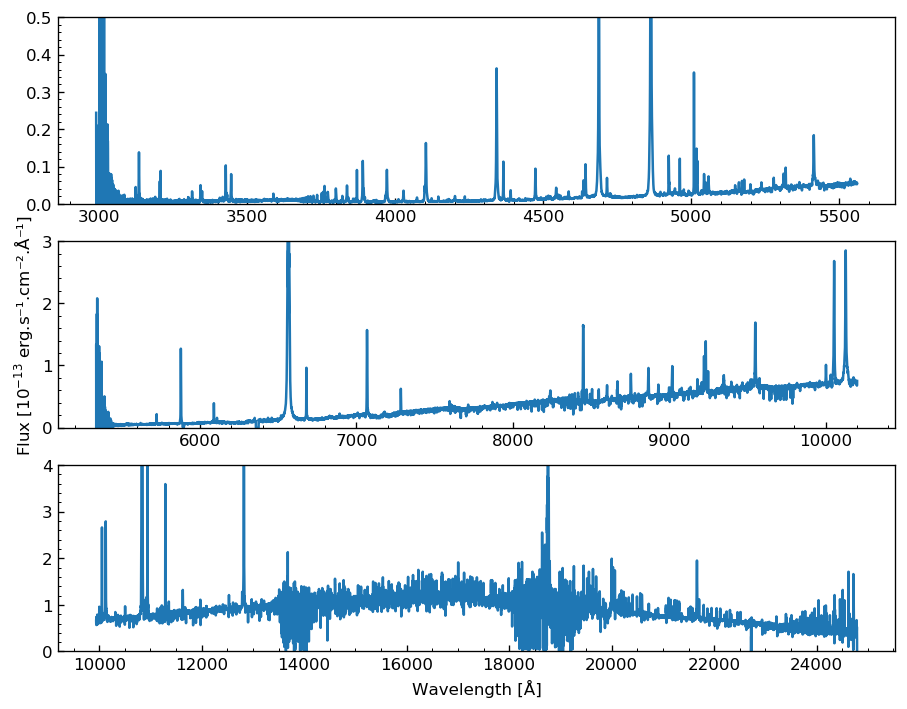}
		\caption{VLT/X-Shooter spectrum of Gaia18aen obtained on March 22, 2018. Upper panel shows spectrum in the UBV arm, middle panel in VIS arm and bottom panel spectrum obtained in NIR arm. The spectrum has been corrected for the telluric features (see the text).}
		\label{fig:VLT_spectrum}
	\end{figure}

	VLT/X-Shooter spectrum of Gaia18aen is shown in Fig. \ref{fig:VLT_spectrum}. A comparison of VLT/X-Shooter spectrum down-sampled to the resolution of the Liverpool Telescope spectrum is shown in Fig. \ref{fig:spectra_comparison}. The identification of the most prominent emission lines visible in the spectra of Gaia18aen is also shown in this figure. Detailed discussion of the spectral features is in Sec. \ref{sec:spectral}.
	
	\subsection{Photometry}
	To study the photometric behavior of Gaia18aen before, during, and after its 2018 outburst, we have collected all available photometric data from the databases of several surveys and from the literature to supplement the light curve in $G$ filter obtained by \textit{Gaia}. Data covering the active stage of Gaia18aen are collected by various telescopes participating in the follow-up network arranged under the Time-Domain work package of the European Commission’s Optical Infrared Coordination Network for Astronomy (OPTICON) grant\footnote{https://www.astro-opticon.org/h2020/network/na4.html}: LCO 0.4-m, PROMPT 0.6-m, Terskol 0.6-m, and PIRATE robotic telescope \citep{2018RTSRE...1..127K}. Data are available in $B$, $V$, $R$, $i$, $g$ filters, and were calibrated using the Cambridge Photometric Calibration Server \citep{2019CoSka..49..125Z,2020arXiv200605160Z}. The calibration process is described in Sec 2.1 of \citet{2020A&A...633A..98W}.
	
	These are supplemented by the data from ASAS-SN survey \citep[$V$ and $g$ filters, ][]{2014ApJ...788...48S, 2017PASP..129j4502K} and with data in $V$ and $I$ from the OGLE IV survey \citep{2015AcA....65....1U} covering mainly the pre-outburst phase (they are saturated during most of the outburst), ATLAS data in non-standard orange and cyan filters \citep{2018PASP..130f4505T}, and by the data from Bochum Survey of the Southern Galactic Disk \citep{2012AN....333..706H, 2015AN....336..590H} in $r$ and $i$ filters. All our photometric data are given in Table \ref{tab:photometry_all} in the Appendix.
	
	Figure \ref{fig:light_curves} shows the individual light curves of Gaia18aen obtained in various filters. We shifted the light curves to the same level for clarity; the values of these shifts are given in the figure legend. In Fig. \ref{fig:outbursts}, part of the light curves showing the active stage of Gaia18aen in 2018 is shown.
	
	\section{Results and discussion}\label{sec:results}
	
	\subsection{Spectral features and symbiotic classification}\label{sec:spectral}
	
\begin{figure}
		\centering
		\includegraphics[width=\columnwidth]{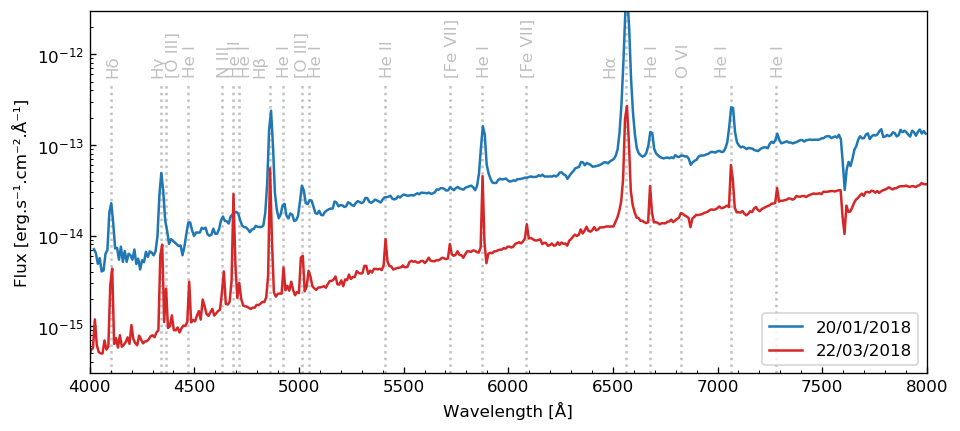}
		\caption{Comparison of the two spectra of Gaia18aen obtained on January 20 and March 22, 2018 together with the identification of the major emission lines observed.}
		\label{fig:spectra_comparison}
	\end{figure}

	Spectroscopic observations of Gaia18aen revealed its symbiotic nature, satisfying the conditions of \citet{2000A&AS..146..407B} - the presence of the absorption features of a late-type giant in the spectrum, and of the emission lines of ions with an ionization potential of at least 35 eV. This classification is further confirmed by the presence of the \ion{O}{vi} in the VLT/X-Shooter spectrum (see Fig. \ref{fig:spectra_comparison}). 
	
	Symbiotic classification of Gaia18aen is also supported by its position in diagnostic diagram using [\ion{O}{iii}] and Balmer lines fluxes and in the \ion{He}{i} diagram \citep[for more details see e.g.][]{2017A&A...606A.110I,2018MNRAS.476.2605I}. In both diagrams \citep[see Fig. 1 and Fig. 4 in][]{2017A&A...606A.110I}, Gaia18aen is located in the region occupied solely by the symbiotic stars ([\ion{O}{iii}] $\lambda$5006/H$\beta$ = 0.04, [\ion{O}{iii}] $\lambda$4363/H$\gamma$ = 0.13; log(\ion{He}{i} $\lambda$6678/\ion{He}{i} $\lambda$5876) = -0.45, log(\ion{He}{i} $\lambda$7065/\ion{He}{i} $\lambda$5876) = -0.18).
	
	In general, Gaia18aen shows an M-star continuum, superimposed with strong emission lines, mainly of \ion{H}{i} and \ion{He}{i}. Comparison of both our spectra shown significant changes in the intensity of the emission lines throughout the outbursts, e.g. significant decrease in case of high ionization lines of \ion{He}{ii}, [\ion{O}{iii}], [\ion{Fe}{vii}] and \ion{O}{vi} (see Sec. \ref{sec:hot}). For clarity of the comparison of both spectra (Fig. \ref{fig:spectra_comparison}), we have down-sampled the spectrum which was obtained by VLT/X-Shooter to the resolution of the spectrum from the Liverpool Telescope (R $\approx$ 350). We have also applied an absolute flux scale to the Liverpool Telescope spectrum using the average $V$ = 12.8 mag such that convolution of the spectrum with the Johnson $V$ filter agrees with the $V$ mag. Table \ref{tab:fluxes} gives the emission-line fluxes that were measured by fitting Gaussian profiles to the calibrated spectra.

	\begin{table}
		\caption{Emission lines fluxes in $\rm 10^{-13}~erg~s^{-1}~cm^{-2}.$}             
		\label{tab:fluxes}      
		\centering                          
		\begin{tabular}{c c c c}        
			\hline\hline                 
			Line &  Wavelength [\AA] & 2018-Jan-20 & 2018-Mar-22 \\
			&   & MJD 58\,139 & MJD 58\,200 \\
			\hline                        
			H$\delta$        & 4\,101.73           & 3.4    & 0.6      \\
			H$\gamma$        & 4\,340.46           & 8.7   & 1.3     \\
			{[}\ion{O}{iii}{]}      & 4\,363.21           & 0.0    & 0.2      \\
			\ion{He}{i}            & 4\,471.48           & 1.4    & 0.2      \\
			\ion{N}{iii}/\ion{N}{iii}*  & 4\,634.13/4\,640.64           & 1.7    &0.1/1.8    \\
			\ion{He}{ii}            & 4\,685.68           & 1.6   & 3.1     \\
			\ion{He}{i}            & 4\,713.15           & 1.8    & 0.1      \\
			H$\beta$         & 4\,861.36           & 45.1  & 6.4    \\
			\ion{He}{i}            & 4\,921.93           & 1.4    & 0.2      \\
			{[}\ion{O}{iii}{]}/\ion{He}{i}*& 5\,006.84/5\,015.68 & 3.5    & 0.3/0.2      \\
			\ion{He}{i}            & 5\,047.74           & 2.1    & <0.1      \\
			\ion{He}{ii}            & 5\,411.52           & 0.6    & 0.5     \\
			{[}\ion{Fe}{vii}{]}     & 5\,721.00           & 0.0    & 0.2      \\
			\ion{He}{i}            & 5\,875.65           & 29.4   & 3.8     \\
			{[}\ion{Fe}{vii}{]}     & 6\,087.00           & 0.0    & 0.4     \\
			H$\alpha$        & 6\,562.80           & 976.9 & saturated \\
			\ion{He}{i}            & 6\,678.15           & 13.8   & 2.3     \\
			Raman \ion{O}{vi}           & 6\,825.00           & 0.0    & 0.5      \\
			\ion{He}{i}            & 7\,065.30           & 51.4  & 5.4    \\
			\ion{He}{i}            & 7\,281.35           & 6.2   & 0.8     \\ 
			\hline                                   
		\end{tabular}
		\tablefoot{* blended lines in the low resolution spectrum from 20/01/2018.}
	\end{table}
	
	\begin{table*}
		\caption{Positions, radial velocities and fluxes of identified emission lines in the VLT/X-Shooter spectrum of Gaia18aen obtained on March 22, 2018. Deredenned fluxes were calculated assuming E(B-V) = 1.2, R$_V$ = 3.1 and extinction law of \citet{1989ApJ...345..245C}. Radial velocities were heliocentric velocity corrected.}             
		\label{tab:fluxes_all}      
		\centering                          
		\begin{tabular}{c c c c c}        
			\hline\hline                 
			Measured $\lambda$ [\AA] & Ion                  & Reference $\lambda$ [\AA]  & Flux [10$^{-14}$ erg s$^{-1}$ cm$^{-2}$]  & Radial velocity [km s$^{-1}$]  \\
			\hline                        
			\dots & \dots &\dots &\dots  &\dots \\
			4974.56 & [\ion{Fe}{vi}] & 4972.47 & 0.1 & 112.2 \\
			4990.38 & [\ion{Fe}{vii}] & 4988.56 & 0.1  & 94.8 \\
			5008.82 & [\ion{O}{iii}] & 5006.85 & 3.3  & 103.2 \\
			5017.49 & \ion{He}{i} & 5015.68 & 2.1 &   93.6 \\
			5020.28 & \ion{Fe}{ii} & 5018.43 & 1.3 &   96.0 \\
			\dots & \dots &\dots &\dots  &\dots \\
			\hline                                   
		\end{tabular}
		\tablefoot{This table is available in electronic format in the full form.}
	\end{table*}
	
	\subsection{Distance and reddening}
	\textit{Gaia} DR2 gives for Gaia18aen a parallax $0.097 \pm 0.054$ mas\,yr$^{-1}$
	with goodness-of-fit statistic parameter \textit{gofAL} $\approx$ 37, that indicates a very poor fit to the data. \citet{2018AJ....156...58B} inferred the distance to the sources from Gaia DR2 using the Bayesian approach which is suitable also for the objects with poor precision of the parallax and even for negative parallaxes. They obtained for Gaia18aen the distance 5.8 kpc placed in the asymmetric confidence interval from 4.6 to 7.6 kpc.
	
	\begin{figure}
		\centering
		\includegraphics[width=\columnwidth]{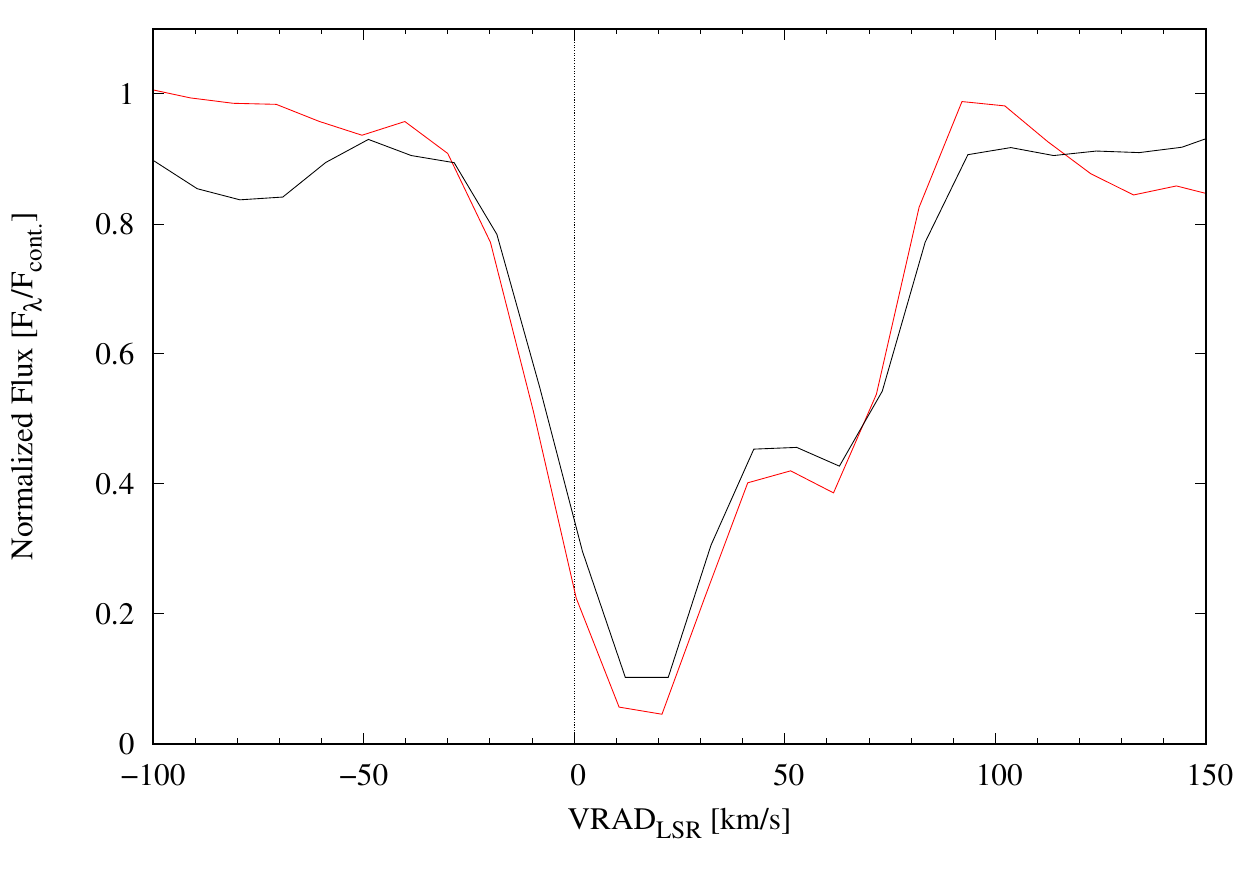}
		\caption{Line profiles of the \ion{Na}{i} D1 (black) and \ion{Na}{i} D2 (red).}
		\label{fig:naI}
	\end{figure}
	
	The \ion{Na}{i} D1, D2 interstellar line profiles reveal two distinct components that allowed us to assume a Doppler splitting caused by at least two diffuse ISM clouds, which may be present along the line-of-sight \citep[see][]{2020arXiv200612877S}. In particular, radial velocities with respect to the local standard of rest, V$_{\rm LSR}$, derived from both \ion{Na}{i} D1, D2 lines indicate a range of mean cloud velocities of $\sim$ 14 to 62 km\,s$^{-1}$ (Fig. \ref{fig:naI}). If the velocity is due to Galactic rotation it can be used to derive a lower limit to distance. In particular, the component with the maximum velocity, V$_{\rm LSR}$ = 62 km\,s$^{-1}$, indicates d > $\sim$ 6 kpc. The radial velocity of the red giant component transformed to LSR, V$_{\rm g,LSR}\sim$81 km\,s$^{-1}$, would indicate a distance of $\sim$ 8 kpc. However, this estimate should be considered as an upper limit because the radial velocity of the giant can be affected by unknown orbital motion and/or pulsation. 
	
	As Gaia18aen is located in the Galactic disc (b = 0.314, Table \ref{tab:gaia}), it is expected to be highly reddened. We estimate total Galactic extinction in its direction, E(B-V) = 1.17  using the maps of the Galactic extinction by \citet{1998ApJ...500..525S}, and E(B-V) = 1.03 from those by \citet{2011ApJ...737..103S}, however, there are important caveats about unreliable extinction estimates for this position.
	
	An independent reddening estimate can be derived from the emission line ratios (Table \ref{tab:fluxes_all}). The \ion{H}{i} Balmer lines are reliable reddening indicators only for negligible self-absorption in the lower series members \citep[case B recombination; e.g.][]{1969ApJ...155..859C, 1975MNRAS.171..395N}. However, this is usually not the case for symbiotic stars \citep[e.g.][]{1997A&A...327..191M} where the reddening-free H$\alpha$/H$\beta$ ratio is $\sim$ 5 - 10 \citep[e.g.][]{1992AJ....103..579M,1995AJ....109.1289M,1994MNRAS.268..213P,1997A&A...327..191M}.  In fact, the values of H$\alpha$:H$\beta$:H$\gamma$:H$\delta$ ratios measured in the spectrum of Gaia18aen are inconsistent with case B recombination for any reddening. Fortunately, we can use \ion{He}{ii} lines which are less prone to optical depth effects. The most useful for the reddening estimate 
	\ion{He}{ii} Pickering line ratios (10124:5411:4542) and Pickering 10124 to Paschen 4686 ratio are consistent with case B and E(B-V) = 1.16 $\pm$ 0.01. Similar value, E(B-V) = 1.14, is indicated by the near IR \ion{H}{i} Br$\gamma$:Pa$\delta$ ratio whereas slightly higher E(B-V) = 1.25 provides better match of the synthetic cool giant spectra to the observed near IR spectrum (see Sec. \ref{sec:cool}).
	
	We have adopted d = 6 kpc and E(B-V) = 1.2 for the rest of this paper.
	
	\subsection{Cool component}\label{sec:cool}
	The VLT/X-Shooter spectrum of Gaia18aen was used to derive atmospheric parameters and the information about the chemical composition of the atmosphere of the cool component. In order to put constraints on the physical parameters of the red giant's atmosphere (mainly its temperature and metallicity) we used the BT-NextGen grid of the theoretical spectra \citet{2011ASPC..448...91A}, calculated using the solar abundances of \citet{2009ARA&A..47..481A} that are available from 'Theoretical spectra webserver' at the SVO Theoretical Model Services\footnote{http://svo2.cab.inta-csic.es/theory/newov2/index.php}. Next, we used the 1D hydrostatic {\sl{MARCS}} model atmospheres by \citet{2008A&A...486..951G} to perform a more detailed analysis of the chemical composition.
	
	We searched among several hundreds of BT-NextGen models in the range of
	parameters: T$_{\rm{eff}}$ = 2\,900 -- 4\,800\,K, $\log{g} = -0.5$ -- 1.5, [M/H] = $-2.5$ -- $+0.5$, and [$\alpha$/H] = $0.0$ -- $+0.4$.  All synthetic
	spectra were convolved with the Gaussian profile with the full width at half
	maximum corresponding to the velocity V = 149\,km\,s$^{-1}$ to achieve the final resolution R$\sim$2\,000. The VLT/X-Shooter spectrum of Gaia18aen was heliocentric velocity corrected by $-14.554$\,km\,s$^{-1}$ and for the red giant velocity, V$_{\rm g, hel}$ = 99 km\,s$^{-1}$, derived from the absorption lines in the near IR spectrum, and then convolved with Gaussian profiles of $v \sin{i}$ = 145\,km\,s$^{-1}$ and $v \sin{i}$ = 147.5\,km\,s$^{-1}$ in the VIS and NIR ranges, respectively, to reduce its resolution to those of adopted for the model spectra. The observed spectrum was dereddened by E$_{\rm{B-V}}$ = 1.25 using \citet{1989ApJ...345..245C} reddening law and adopting total to selective absorption ratio R = 3.1 with use of {\sl VOSpec}\footnote{https://www.cosmos.esa.int/web/esdc/vospec} Virtual
	Observatory tool. The adopted initial value E$_{\rm{B-V}}$ = 1.20 was
	replaced with E$_{\rm{B-V}}$ = 1.25 to achieve better compliance in the
	$J$-band region.
	
	Each spectrum was normalized with the flux value measured in a narrow (50\,\AA) range of the $K$-band region centered at 22\,155\,\AA.  The residuals were calculated for each pair of spectra (theoretical model and the observed spectrum) to obtain the $\chi^2$ value that characterizes the fit quality.  Only selected ranges in the NIR of the $H$- and $K$-band regions were finally used in the residuals calculations to exclude the areas disturbed by some artifacts. Shorter wavelength region was not taken into account as in the case of symbiotic systems the visual range is strongly dominated by the contribution from the hot component and the nebula, and even in the $J$-band, there are present numerous, strong emission lines (see Table \ref{tab:fluxes_all}).
	
	\begin{figure}
		\centering
		\includegraphics[width=\columnwidth]{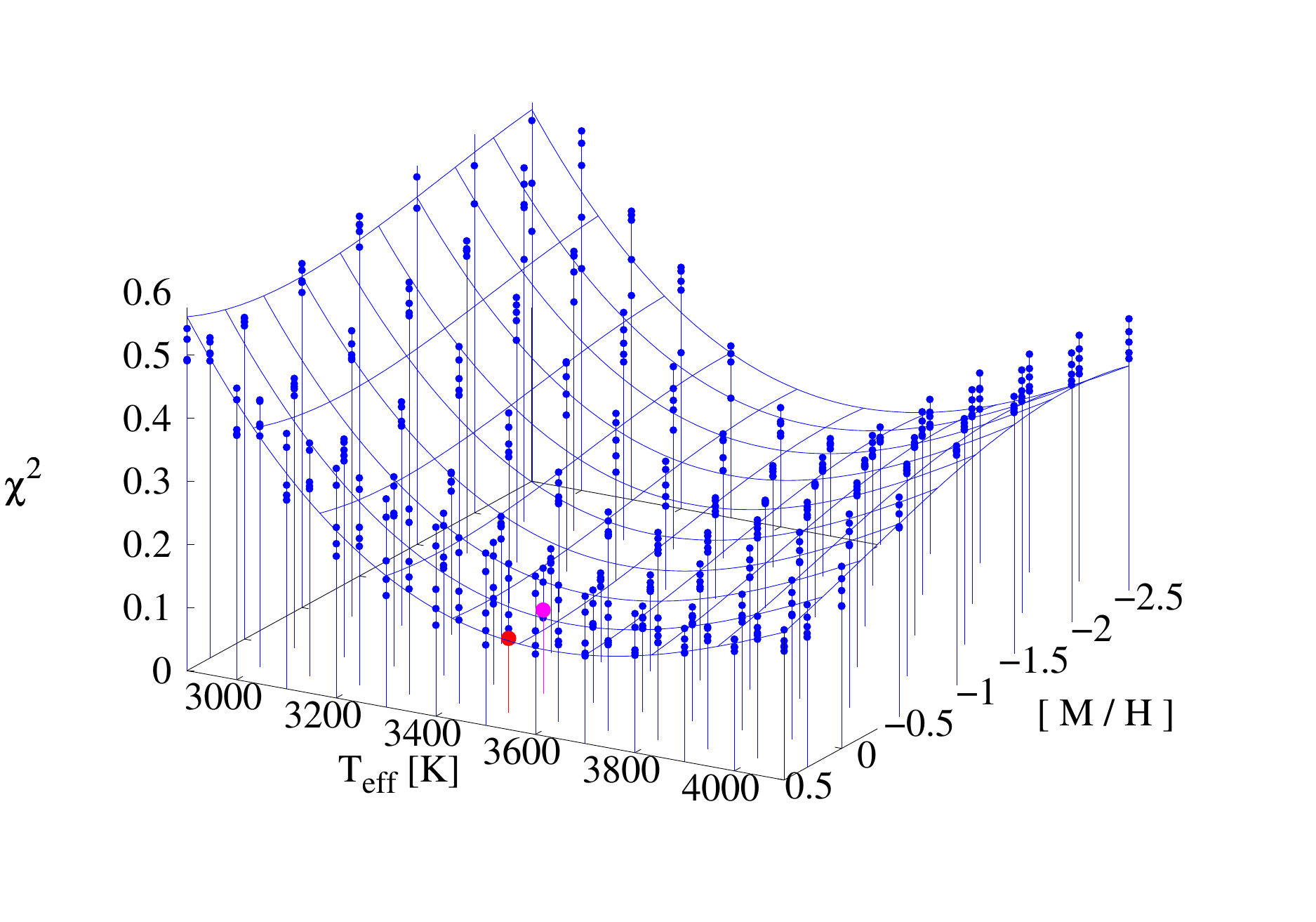}
		\caption{$\chi^2$ surface in the function of T$_{\rm{eff}}$ and [M/H]
			parameters. Two selected cases are distinguished with red (the best match)
			and magenta (with atmospheric parameters close to those obtained via
			spectral synthesis fit -- see Fig. \ref{fig:sed} and \ref{fig:model_kband}).}
		\label{fig:3dfit}
	\end{figure}
	
\begin{figure}
		\centering
		\includegraphics[width=\columnwidth]{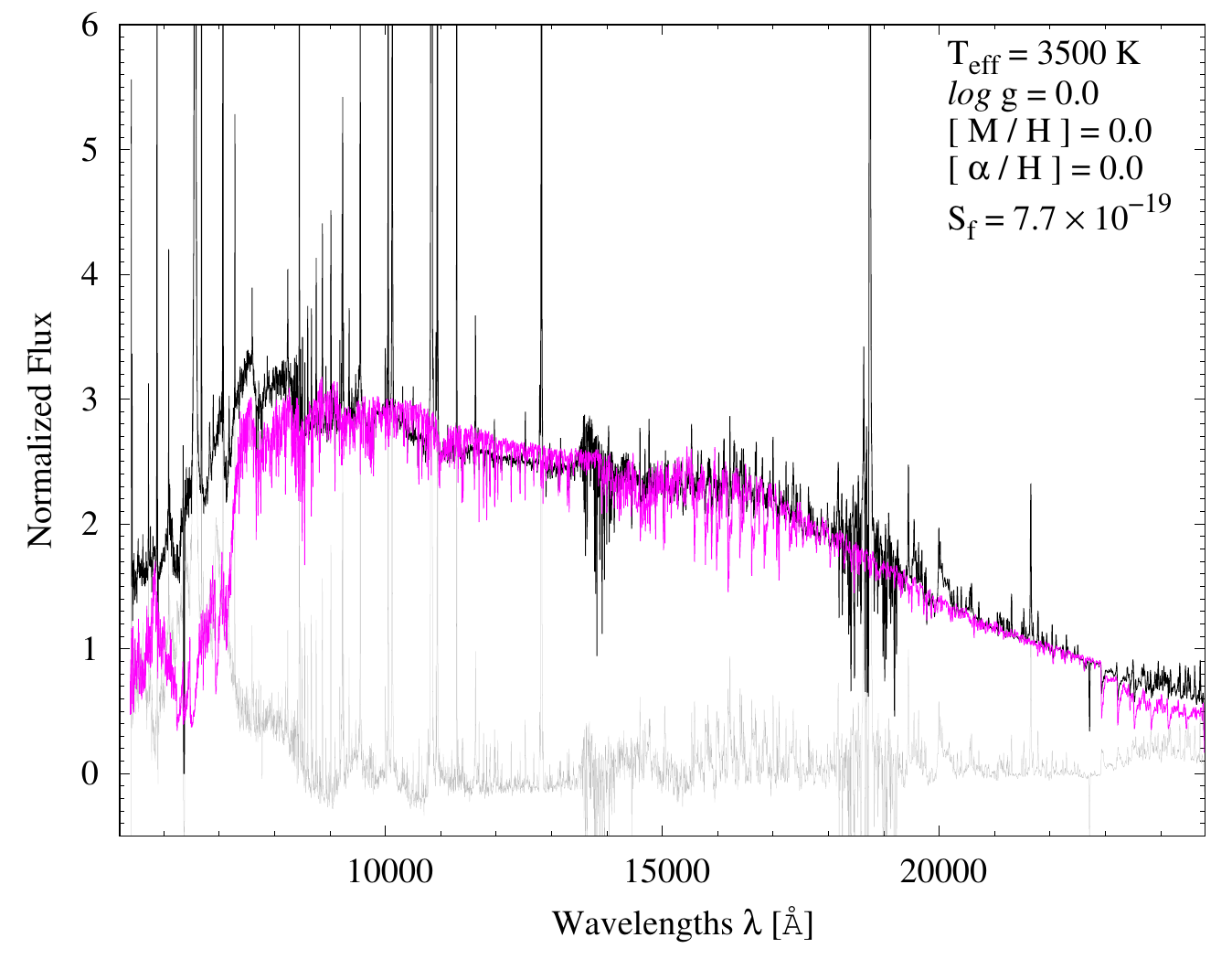}
		\caption{Comparison of the VLT/X-Shooter spectrum and the synthetic model. The VLT/X-Shooter spectrum (black line) dereddened by E$_{\rm{B-V}}$ = 1.25 is compared with the synthetic model (magenta) corresponding to
			atmospheric parameters: T$_{\rm{eff}}$ = 3500\,K, $\log{g} = 0.0$, [M/H] =
			$0.0$, and [$\alpha$/H] = $0.0$.  The residuals - observations minus
			calculations - are shown at the bottom with gray line. The spectra were
			normalized to the flux level at the narrow range ($50$\,\AA) centered at
			$22\,155$\,\AA.  The scale factor is $S_f = F_{obs}/F_{model} = 7.7 \times
			10^{-19}$.}
		\label{fig:sed}
	\end{figure}

	The fit quality in the parameter space is illustrated in Fig. \ref{fig:3dfit}. We obtain the strongest dependence on temperature. The best solutions are obtained for T$_{\rm{eff}} \sim$ 3\,500\,K.  In the case of the remaining parameters, there is significant degeneration and the minimum is not clearly defined, however, in the case of metallicity the near solar values seem to be preferred. Taking into account the values of the scale factor (generally in the range $S_f =$ 6.4 -- 7.9 $\times 10^{-19}$ in the case of the best matching models), we can limit the value of surface gravity $\log{g} \approx 0.0$ as the higher value could result in a too high mass of the giant. Fig. \ref{fig:sed} shows the comparison of the VLT/X-Shooter spectrum with the synthetic model corresponding to the atmospheric parameters T$_{\rm{eff}}$ = 3\,500\,K, $\log{g} = 0.0$, [M/H] = $0.0$, and [$\alpha$/H] = $0.0$ close to those finally preferred from the spectral synthesis.
	
\begin{figure*}
		\centering
		\resizebox{\hsize}{!}
		{\includegraphics[width=\columnwidth]{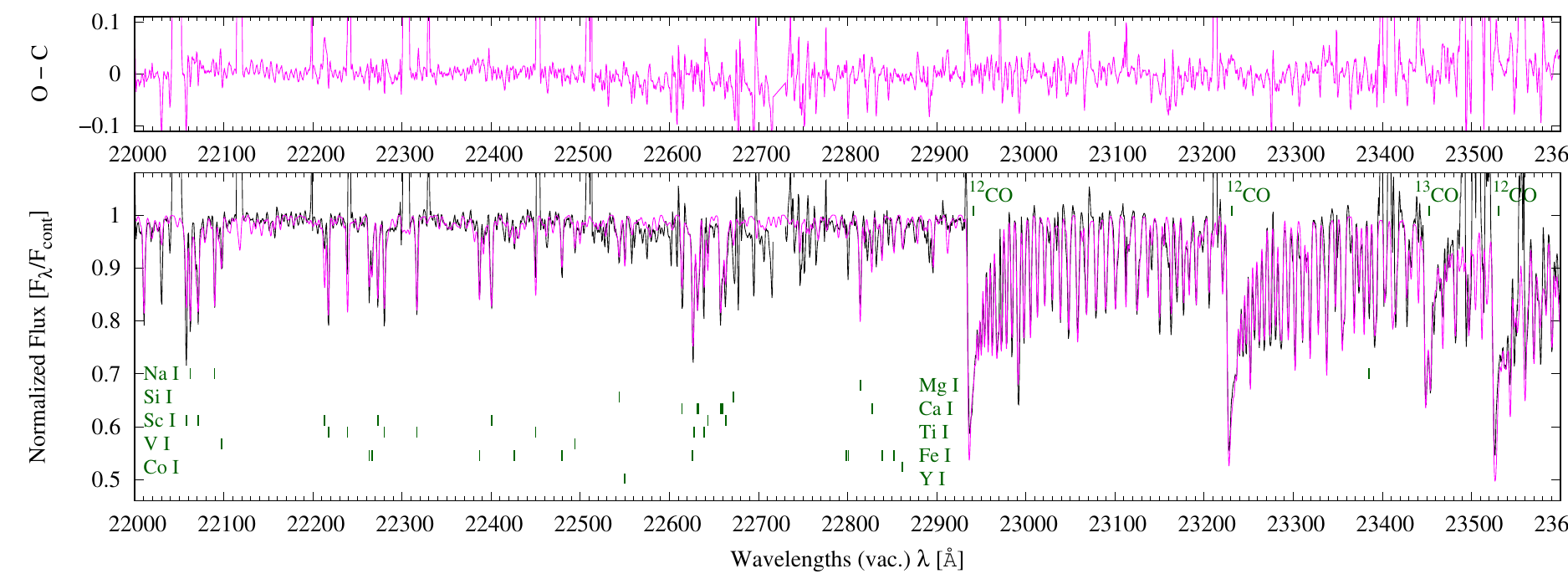}}
		\caption{VLT/X-Shooter spectrum of Gaia18aen (black line) and synthetic spectrum (magenta line) calculated using the final abundances (Table \ref{tab:abundances}).}
		\label{fig:model_kband}
	\end{figure*}

	The elemental abundances of the particular elements were measured through the fit of the synthetic spectrum to the observed one in the $K$-band region. The observed spectrum has been normalized beforehand to the continuum level. We tested a number of the atmosphere models \citep[MARCS;][]{2008A&A...486..951G} with temperature and surface gravity set to constant values as follows: T$_{\rm{eff}}$ = 3\,500\,K, $\log{g} = 0.0$. Various metallicities expressed in [Fe/H] were tested in the range $-0.5$ -- $+0.5$\,dex including a sample of alpha enhanced cases ([$\alpha$/Fe] $= +0.4$\,dex). The micro- ($\xi_{\rm t}$) and macro- ($\zeta_{\rm t}$) turbulence velocities, were set to 2\,km\,s$^{-1}$ and 3\,km\,s$^{-1}$, respectively -- the values typical for cool, Galactic red giants. The excitation potentials and $gf$ values for transitions in the case of atomic lines were taken from the Vienna Atomic Line Database \citep{1999A&AS..138..119K}. For the molecular data, we used line lists by \citet{1994ApJS...95..535G} for CO, R. L. Kurucz\footnote{http://kurucz.harvard.edu} for OH, and \citet{2014ApJS..214...26S} for CN. The spectrum synthesis was run using the {\sl{WIDMO}} code \citep{2006A&A...446..603S} with the method as described by \citet[][and ref. therein]{2017MNRAS.466.2194G}. The best match was obtained for the model with slightly super-solar metallicity [Fe/H] $= +0.25$\,dex and [$\alpha$/Fe] $= 0.0$\,dex with differences of abundances in relation to the model not larger than 0.11\,dex in the case of elements that are best represented with atomic lines in the spectrum: Fe, Ti, Ca, and Na. The obtained abundances are listed in Table \ref{tab:abundances} and the synthetic fit to the observed spectrum is shown in Fig. \ref{fig:model_kband}.
	
	\begin{table*}
		\caption{Final values of abundances obtained from $K$-band region
			together with the formal fitting errors, and $^{12}$C/$^{13}$C isotopic
			ratio.  The results for the two best-matching {\sl{MARCS}} models with the
			same T$_{\rm{eff}}$ = 3\,500\,K, $\log{g} = 0.0$, $[$Fe/H$]$ $= +0.25$\,dex
			and different abundances of the $\alpha$ elements are presented on the left
			($[\alpha$/Fe$]$ $= 0.0$\,dex), and on the right ($[\alpha$/Fe$]$ $=
			+0.4$\,dex).}             
		\label{tab:abundances}      
		\centering          
		\begin{tabular}{@{}l@{\hskip 5mm}llll@{\hskip 10mm}l@{}}
			\hline\hline
			Model:            & \multicolumn2c{$[\alpha$/Fe$]$=$0.0$} & \multicolumn2c{$[\alpha$/Fe$]$=$+0.4$} & \\
			\hline
			Element           & $\log\varepsilon$($X$)$^a$ & [$X$]$^b$ & $\log\varepsilon$($X$)$^a$ & [$X$]$^b$ & Unit\\
			\hline      
			C                 & $7.40\pm0.13$ & $-1.03\pm0.18$ & $7.43\pm0.07$ & $-1.00\pm0.12$ & dex \\
			N                 & $8.92\pm0.35$ & $+1.09\pm0.40$ & $9.52\pm0.33$ & $+1.69\pm0.38$ & dex \\
			O                 & $8.03\pm0.29$ & $-0.66\pm0.34$ & $8.75\pm0.28$ & $+0.06\pm0.33$ & dex \\
			Na                & $6.52\pm0.19$ & $+0.31\pm0.23$ & $6.64\pm0.24$ & $+0.43\pm0.28$ & dex \\ 
			Mg                & $8.65\pm0.13$ & $+1.06\pm0.17$ & $8.70\pm0.08$ & $+1.11\pm0.12$ & dex \\ 
			Si                & $8.29\pm0.36$ & $+0.78\pm0.39$ & $8.12\pm0.37$ & $+0.61\pm0.40$ & dex \\ 
			Ca                & $6.59\pm0.13$ & $+0.27\pm0.16$ & $6.69\pm0.19$ & $+0.37\pm0.22$ & dex \\ 
			Sc                & $3.67\pm0.18$ & $+0.51\pm0.22$ & $3.69\pm0.11$ & $+0.53\pm0.15$ & dex \\ 
			Ti                & $5.23\pm0.15$ & $+0.30\pm0.19$ & $5.34\pm0.15$ & $+0.41\pm0.19$ & dex \\ 
			V                 & $3.76\pm0.35$ & $-0.13\pm0.43$ & $3.72\pm0.27$ & $-0.17\pm0.35$ & dex \\ 
			Fe                & $7.81\pm0.18$ & $+0.34\pm0.22$ & $7.68\pm0.14$ & $+0.21\pm0.18$ & dex \\ 
			Co                & $5.23\pm0.36$ & $+0.30\pm0.41$ & $5.19\pm0.54$ & $+0.26\pm0.59$ & dex \\ 
			Y                 & $2.28\pm0.28$ & $+0.07\pm0.33$ & $1.93\pm0.23$ & $-0.28\pm0.28$ & dex \\ 
			$^{12}$C/$^{13}$C & \multicolumn2c{$10.0\pm1.5$}   & \multicolumn2c{$9.7\pm1.6$}    & 1   \\
			\hline                  
		\end{tabular}
		\tablefoot{$^a$ $\log{\epsilon}(X) = \log{(N(X) N(H)^{-1})} + 12.0$. Uncertainty is in 3\,$\sigma$.\\$^b$ Relative to the Sun [$X$] abundances in
			relation to the solar composition of \citet{2009ARA&A..47..481A}; \citet{2015A&A...573A..26S,2015A&A...573A..25S}.}
	\end{table*}
	
	The relative [Ti/Fe] = -0.04 abundance compared to metallicity ([Fe/H] = 0.34) is consistent with the membership in Galactic disk \citep[see][fig. 7]{2017MNRAS.466.2194G} whereas [Ti/Fe] ratio in relation to [Ti/H]
	suggests that Gaia18aen belongs to the disk population with the age\,$<
	7$\,Gyr \citep[see][]{2014A&A...562A..71B}. Practically all derived abundances in relation to iron ([Na/Fe], [Ca/Fe], [Ti/Fe], [Y/Fe]) are consistent with those expected for the disk population \citep{2014A&A...562A..71B}. The only exception is [Mg/Fe] and [Si/Fe] that seems to indicate a large overabundance of these two elements. However, these abundances are less reliable, as they are based on a few weak lines in the case of \ion{Si}{i} or a single relatively strong feature, which is split into the number of lines in the case of \ion{Mg}{i}. Our best solution for the model with [$\alpha$/Fe] $= 0.0$\,dex resulted with suspiciously low oxygen abundance ([O/H] $= -0.66\pm0.34$), that is difficult to explain by the models of evolution in the symbiotic systems. The second best-fit model obtained for the same atmospheric parameters and enhanced [$\alpha$/Fe] $= 0.4$\,dex gives much more reliable value [O/H] $= +0.06\pm0.33$ (Table \ref{tab:abundances}). However, the abundances of oxygen and nitrogen as well as some other elements derived from atomic lines (e.g. Si, V, Co) can be burdened with large errors that can account for some peculiarities of the obtained composition.
	
	\citet{2019Sci...365..478S} have mapped the shape of the Milky Way disk based on the distances to nearly 2\,500 classical Cepheids. In their coordinate system (R, $\phi$; where R is the distance of the object from Galactic center and $\phi$ is the Galactocentric azimuth measured counterclockwise from
	$l=0^\circ$), Gaia18aen would be placed at R$\sim$11.5\,kpc and $\phi
	\sim 28^\circ$. At this location, the disk is bent south, however, the displacement of Gaia18aen away from the central disk surface is only $\sim 0.2$\,kpc north that is in agreement with its disk membership.
	
	The near IR spectrum as well as 2MASS and $WISE$ colors are all consistent with a non-dusty S-type symbiotic system. The scaling factor resulting from our best model fit combined with the distance to Gaia18aen implies the red giant radius of $\sim 230\,R_\sun$ and luminosity of $\sim 7\,400\,L_\sun$ that places Gaia18aen among the brightest symbiotic giants \citep[e.g.][]{2012BaltA..21....5M}. 
	
	The giants in S-type symbiotic systems often exhibit pulsations and moreover, S-type systems have orbital periods on such timescales \citep[$\sim$ 300 - 600 days;][]{2013AcA....63..405G}, which should be detectable using the available data. We have therefore used out-of-outburst data (OGLE $I$ and Gaia $G$, with data from the period JD $\sim$ 2\,458\,000 - 2\,458\,500 excluded and ATLAS $o$ and $c$, Bochum $r$ and $i$) to search for any periodical changes. There are several clear minima in these quiescent light curves, around JD $\sim$ 2\,455\,910, 2\,456\,355, 2\,457\,345 and 2\,458\,820, respectively. The periodicity that fits this data is:
	
	\begin{equation}
		{\rm Min} = (2\,457\,349.4 \pm 8.3) + (486.9 \pm 3.9) \times E
	\end{equation}
	
	ASAS-SN light curves were not used in this analysis, as their coverage is insufficient (especially in $g$ filter) and/or their scatter is larger than the amplitude in the best dataset (combined ATLAS $o$ and Gaia $G$). The phased light curve in $g$, however, seems to show the same modulation whereas it is less obvious in $V$. The phased light curve in selected filters is shown in Fig. \ref{fig:phased_light_curves}.
	
	We tentatively attributed the 487-day modulation to the orbital period of the system. Assuming the total mass of the system is $\sim$\,2 - 3\,M$_\sun$ \citep[i. e. similar to other symbiotic stars - see e.g.][]{2003ASPC..303....9M} we estimate the binary separation to be about 1.5 - 1.7\,AU. The red giant with a radius of $\sim$\,230\,R$_\sun$ could then fill its Roche lobe. While there is some indication for a secondary minimum due to ellipsoidal variability in the phased light curve (Fig. \ref{fig:phased_light_curves}), a well-sampled, long-term light curve of Gaia18aen and/or measurements of giant's radial velocities would be needed, to fully confirm this finding and to refine the period.
	
	The large scatter in quiescent light curves may be due to additional short-term variations with timescales of 50 - 200 days caused by stellar pulsations of the red giant component of the binary system. It can be either a semi-regular variable or so-called OSARG (OGLE Small Amplitude Red Giant). Thus, the red giant in Gaia18aen would be very similar to red giants in S-type symbiotic systems - also from this point of view \citep[see e.g.][for more details about light curves of S-type symbiotics]{2013AcA....63..405G}. 
	
	\begin{figure*}
		\centering
		\resizebox{\hsize}{!}
		{\includegraphics[width=\columnwidth]{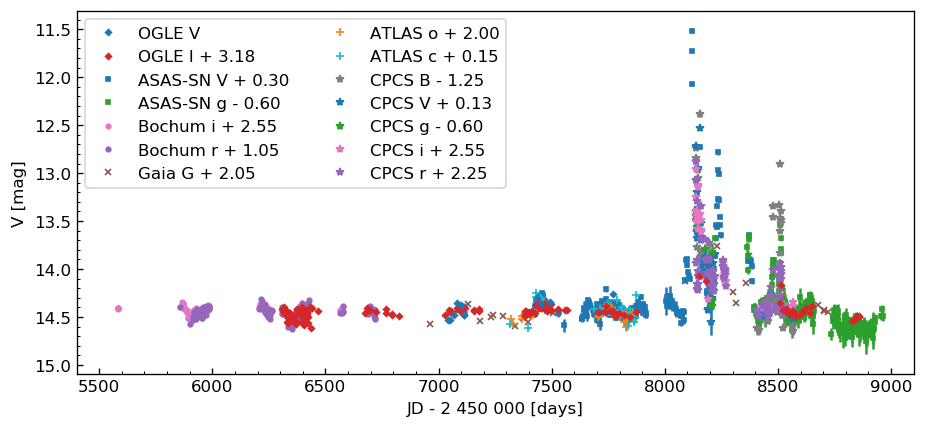}}
		\caption{Light curves of Gaia18aen. Individual light curves in various filters were shifted to the level in the OGLE $V$ filter for clarity, values of shifts are shown in the figure legend. Different colors denote different filters.}
		\label{fig:light_curves}
	\end{figure*}
	
	\begin{figure}
		\centering
		\includegraphics[width=\columnwidth]{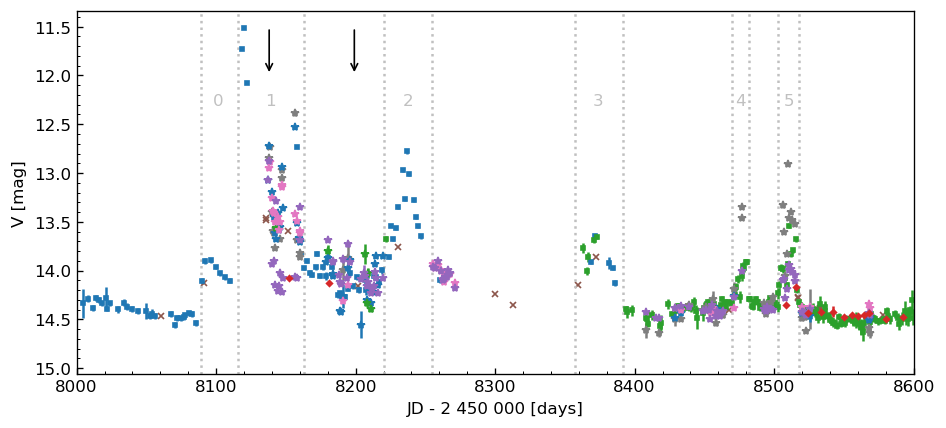}
		\caption{Outbursts of Gaia18aen. The color are same as in Fig. \ref{fig:light_curves}. Vertical dotted lines indicate the individual brightenings (denoted by numbers 0 - 5), described in the text in more details. The arrows show the times when the two spectra were obtained.}
		\label{fig:outbursts}
	\end{figure}
	\begin{figure}
		\centering
		\includegraphics[width=\columnwidth]{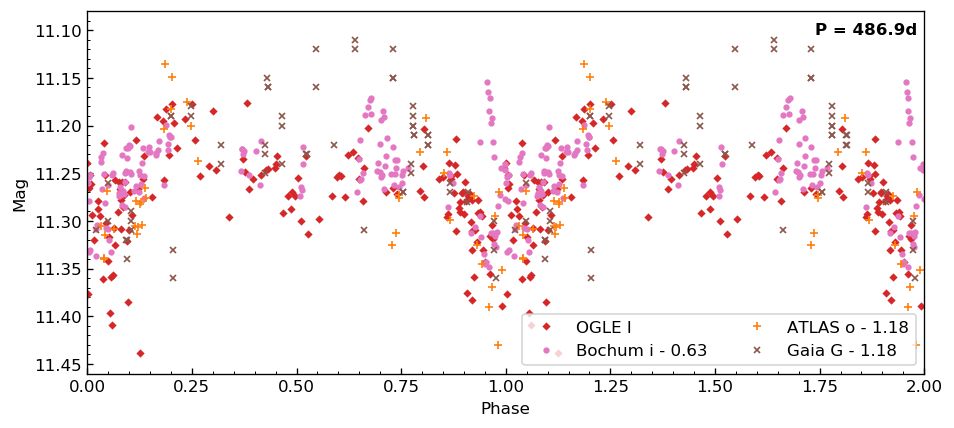}
		\caption{Light curves in selected filters phased with the period of P = 486.9 days. Individual data-sets were shifted to the level in the OGLE $I$ filter.}
		\label{fig:phased_light_curves}
	\end{figure}
	
	\subsection{Hot component and outburst activity}\label{sec:hot}
	The most prominent features in the light curves of Gaia18aen (Fig. \ref{fig:light_curves}) are the outbursts observed in 2018. In Fig. \ref{fig:outbursts}, the part of the light curve showing the active stage is depicted. The light curves in various filters were shifted to the same level for clarity, in order to study the structure of the active stage. Individual brightenings are labeled by numbers 0 - 5. 
	
	First small pre-outburst of about 0.5 mag in $V$ filter (denoted 0) was detected at JD 2\,458\,089. It was followed by more prominent outburst (1) approximately after 27 days (JD 2\,458\,116) of gradual decline. The maximal brightness during this outburst was $\approx$ 11.2 mag in $V$ filter (JD 2\,458\,119), thus the amplitude was about 3.3 mag in comparison with the quiescent V magnitude (OGLE V $\sim$ 14.5). The analysis of the combined light curves constructed from the observations in various filters revealed complex structure of this outburst with at least two another re-brightnenings which followed the first maximum. Approximately after 47 days (JD 2\,458\,163) after the first maximum Gaia18aen reached the magnitude similar to the brightness of the pre-outburst 0. Outburst denoted in this paper as 2 started after 57 days of a quasi-steady period (JD 2\,458\,220) and lasted for 35 days (until JD 2\,458\,255). The maximal brightness in $V$ filter during this period was $\approx$ 12.5 mag.
	
	Third, small scale outburst was detected at JD 2\,458\,357. The amplitude of the outburst was much smaller than in the previous cases, however, the approximate duration was similar to the outburst 2. Up-to-now, the last two increases in brightness (denoted 4 and 5) have been detected at JD 2\,458\,470 and JD 2\,458\,503. It is worth noting that both brightenings were much shorter than the previous ones, 12 and 15 days, respectively. Moreover, the shape of at least the first one (4) looks suddenly interrupted which may indicate that both brightenings are part of a single outburst that is apparently interrupted by some obscuration process. Since JD 2\,458\,518, there were no other significant brightenings detected in the case of Gaia18aen. The amplitude of the outbursts and their duration resemble the behavior of typical classical symbiotic stars \citep[e.g. AG Dra, Z And; see e.g.][]{1995AJ....109.1289M,1996AJ....112.1659M,2019CoSka..49..228M,2019arXiv190901389M}. Multiple outbursts with timescales similar to those observed in Gaia18aen are predicted also by some nova models \citep{2014MNRAS.437.1962H}. 
	
	In addition to photometric evolution of the brightenings, we have analyzed two spectra of Gaia18aen, obtained during its activity. The first spectrum was obtained during a decline from the outburst 1, 20 days after the optical maximum when the optical brightness dropped by about 1.5 mag whereas the second one was obtained 81 days after this maximum when the optical brightness was in the period of low brightness ($V$ $\sim$ 14.0) in the middle between the first outburst and the re-brightening observed approximately after 100 days (see the arrows in Fig. \ref{fig:outbursts} which are showing the times when the spectra were obtained).
	
	The comparison of the obtained spectra of Gaia18aen is shown in Fig. \ref{fig:spectra_comparison}. In Table \ref{tab:fluxes}, we present the measured absolute fluxes of the most prominent emission lines detected in the optical part of the spectra. It is clear from the comparison, that the outburst activity of Gaia18aen was accompanied by the significant changes in its spectra. In general, the fluxes of emission lines of \ion{H}{i} and \ion{He}{i} decreased by a factor of $\sim$\,8, between the time when the first spectrum was obtained and the second spectrum obtained in the period between the outbursts, especially as a result of the decreasing continuum. In the same time, fluxes of high ionization lines are either much lower in the first spectrum of Gaia18aen (\ion{He}{ii}), or the lines are even not detectable ([\ion{O}{iii}], [\ion{Fe}{vii}], and \ion{O}{vi}). Such behavior indicates increasing ionization as the system declines from the outburst maximum, similar to those observed during symbiotic star outbursts \citep[e.g.][]{1991AJ....101..637K,1992AJ....103..579M,1995AJ....109.1289M,1999A&A...347..478G,2016MNRAS.456.2558L, 2019CoSka..49..228M}.
	
	The maximum optical magnitude recorded for Gaia18aen was $V$ = 11.2 mag on JD 2\,458\,119. Assuming that most of the hot component continuum emission is shifted to the optical (a lower limit to L if not) and that during outburst m$\rm _{bol}$ $\sim$ $V_{hot}$, we estimate the reddening corrected $m_{\rm bol , 0} \approx 7.5$, and the absolute bolometric magnitude M$\rm _{bol}$ $\sim$ -6.4 mag which corresponds to L $\sim$ 28\,000 L$_\sun$. This approach assumes that in case of Gaia18aen - like in other symbiotic stars - at the strong outburst maximum the optical bands are dominated by the A or F-type photosphere (T$_{\rm{eff}}$ $\sim$ 9\,000\,K), rather than a nebular continuum that is confirmed by spectroscopic observations \citep[e.g.][]{1986syst.book.....K,1992AJ....103..579M}.
	
	To estimate the temperature and luminosity of the hot component during the decline, we make use of the emission line fluxes. The minimum temperature is set by the maximum ionization potential (IP) observed in the spectrum, and the relation T [10$^3$ K] $\sim$ IP$_{\text{max}}$ [eV] found by \citet{1994A&A...282..586M}, to give T $\sim$ 55 kK, and 114 kK, using the spectra obtained 20 and 81 days after the optical maximum, respectively. An upper limits for T of 80 kK and 155 kK for the first and second epoch, respectively, were derived from \ion{He}{ii} $\lambda$4\,868, \ion{He}{i} $\lambda$5\,876 and H$\beta$ emission line ratios assuming Case B recombination \citep{1981psbs.conf..517I}. We estimate the luminosity of the hot component to be L $\sim$ 21\,000 L$_\sun$ and 5\,200 L$_\sun$, for the first and second epoch, respectively, using equation 8 of \citet{1991AJ....101..637K}. Similarly, equations 6 and 7 of \citet{1997A&A...327..191M} give L(\ion{He}{ii} $\lambda$4\,868) $\sim$ 29\,600 L$_\sun$ and 5\,890 L$_\sun$, and L(H$\beta$) $\sim$ 29\,200 L$_\sun$ and 5\,340 L$_\sun$ for the two epochs, respectively.
	
	All these estimates assume a black-body spectrum for the hot component and case B recombination for the emission lines and are accurate to only a factor of $\sim$ 2. They also assume d = 6 kpc and E(B-V) = 1.2. 
	
\begin{figure}
		\centering
		\includegraphics[width=\columnwidth]{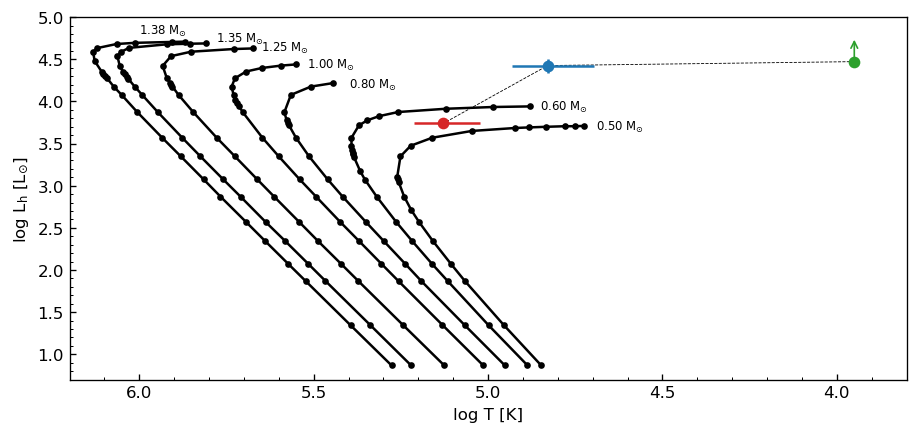}
		\caption{Evolution of the hot component of Gaia18aen in the HR diagram throughout its outburst. Green symbol corresponds to the optical maximum, blue and orange symbols represent the values calculated using the datasets obtained 20 and 81 days after the outburst maximum, respectively. The solid curves with dots are steady models of \citet{2007ApJ...663.1269N}.}
		\label{fig:HR}
	\end{figure}

	The evolution of the hot component in the HR diagram is shown in Fig. \ref{fig:HR}. During the first phase of the main outburst, the hot component luminosity remained almost constant (at least from the optical maximum until our spectrum obtained 21 days after maximum). The temperature increased $\sim$ 10 times in the same period. This was later followed by a slight increase in temperature and a decline of luminosity. The outbursts of the classical symbiotic stars such as Z And, CI Cyg, or AX Per are typically accompanied by the decrease in the temperature during the optical maxima, while the luminosity remains roughly constant throughout the outburst \citep[see e.g. Fig. 1 in][]{2010arXiv1011.5657M}. On the other hand, in case of AG Dra, luminosity may increase by a factor of 5 -- 10 throughout its outburst \citep{1995AJ....109.1289M}, that is similar to the case of Gaia18aen, when we compare the maximum phase with the period 81 days after maximum.
	
	The high luminosity and temperature indicate that Gaia18aen could be detectable in soft X-rays. On the other hand, the object is highly reddened and probably therefore the source has not been reported in X-rays yet.
	
	\section{Conclusions}
	In this work, we have analyzed the photometric and spectroscopic observations of Gaia18aen, transient detected by the \textit{Gaia} satellite at the beginning of the year 2018. Our main findings are as follows:
	\begin{itemize}
		\item Gaia18aen is a classical symbiotic star, fulfilling the traditional criteria for symbiotic stars. Raman scattered \ion{O}{vi} lines are observed in its spectra outside the outbursts.
		\item The system is located at the distance $\sim$\,6\,kpc, 0.2\,kpc from the central disk surface.
		\item The cool component of this symbiotic binary is an M giant with T$_{\rm eff}$\,$\sim$\,3\,500\,K of a slightly super-solar metallicity [Fe/H] $= +0.25$\,dex, with a radius of $\sim 230\,R_\sun$. Its luminosity, L\,$\sim\,7\,400\,L_\sun$, makes it one of the brightest symbiotic giants. The near IR spectrum and IR photometry from 2MASS and WISE are consistent with a non-dusty S-type symbiotic star. 
		\item The system experienced an outburst of about 3.3\,mag in 2018, followed by re-brightening detected approximately after 100, 240, and 350\,days. At least the first outburst was accompanied by the increase of the hot component luminosity (L$_h$ $\sim$ 28\,000\,L$\sun$ at the optical maximum) and the decrease in temperature (A or F-type photosphere), in comparison with temperature $\sim$\,68\,kK and $\sim$\,135\,kK, and luminosity of $\sim$ 26\,600 and 5\,500\,L$\sun$, corresponding to the observations obtained 20 and 81 days after the optical maximum, respectively.
		\item The outburst was accompanied by the changes in emission spectral lines typical for classical symbiotic stars. In outburst, higher fluxes of lower ionization lines of \ion{H}{i} and \ion{He}{i} have been observed, together with the decrease of intensity of high ionization lines of \ion{He}{ii}, [\ion{O}{iii}], [\ion{Fe}{vii}], and \ion{O}{vi}.
		\item The quiescent light curves of the object are characterized by a periodicity of approximately 487\,days, which we tentatively attributed to the orbital modulation. The scatter in the light curves might be caused by stellar pulsations of the red giant with a period of 50 - 200\,days, which are typical for cool components in S-type symbiotics.
	\end{itemize}
	
	These findings make Gaia18aen the first symbiotic star discovered by the \textit{Gaia} satellite.

	\begin{acknowledgements}
		The research of JaM was supported by the Charles University, project GA UK No. 890120 and by internal grant VVGS-PF-2019-1047 of
		the Faculty of Science, P. J. \v{S}af\'{a}rik University in Ko\v{s}ice. This research has been partly founded by the National Science Centre, Poland, through grant OPUS 2017/27/B/ST9/01940 to JM. MG and JS are supported by the Polish NCN MAESTRO grant 2014/14/A/ST9/00121. CG has been financed by Polish NCN grant SONATA No. DEC-2015/19/D/ST9/02974.
		{\L}W acknowledges support from the Polish NCN grant Daina No. 2017/27/L/ST9/03221. {\L}W and PZ acknowledge support from EC Horizon 2020 grant No 730890 (OPTICON). 
		
		The Faulkes Telescope Project is an education partner of Las Cumbres Observatory (LCO). The Faulkes Telescopes are maintained and operated by LCO. We thank David Asher of Armagh Observatory for collecting part of the data.
		
		We acknowledge the use of the Cambridge Photometric Calibration Server (http://gsaweb.ast.cam.ac.uk/followup), developed by Sergey Koposov and maintained by {\L}ukasz Wyrzykowski, Arancha Delgado, Pawe{\l} Zieli{\'n}ski, funded by the European Union's Horizon 2020 research and innovation programme under grant agreement No 730890 (OPTICON).
		
		This work has made use of data from the European Space Agency (ESA) mission {\it Gaia} (\url{https://www.cosmos.esa.int/gaia}), processed by the {\it Gaia} Data Processing and Analysis Consortium (DPAC, \url{https://www.cosmos.esa.int/web/gaia/dpac/consortium}). Funding for the DPAC
		has been provided by national institutions, in particular the institutions participating in the {\it Gaia} Multilateral Agreement; the use of the SIMBAD and VIZIER databases, operated at CDS, Strasbourg, France; the data products from the Two Micron All Sky Survey, which is a joint project of the University of Massachusetts and the Infrared Processing and Analysis Center/California Institute of Technology, funded by the National Aeronautics and Space Administration and the National Science Foundation and the data products from the Wide-field Infrared Survey Explorer, which is a joint project of the University of California, Los Angeles, and the Jet Propulsion Laboratory/California Institute of Technology, funded by the National Aeronautics and Space Administration.
	\end{acknowledgements}
	
	%
	\bibliographystyle{aa} 
	\bibliography{paper.bib} 
	%
	
	\begin{appendix} 
		\section{Photometrical observation}
		\begin{table}[h!]
			\caption{Photometrical observations of Gaia18aen.}             
			\label{tab:photometry_all}      
			\centering          
			\begin{tabular}{c c c c c}     
				\hline\hline       
				JD 24.. & Mag & Magerr & Filter & Source \\\hline   
				\dots & \dots & \dots & \dots & \dots \\
				57840.54 & 14.18 & 0.03 & $V$ & ASAS-SN \\
				57843.31 & 14.35 & 0.02 & $c$ & ATLAS \\
				57843.33 & 14.33 & 0.02 & $c$ & ATLAS \\
				57843.34 & 14.36 & 0.01 & $c$ & ATLAS \\
				57843.38 & 14.38 & 0.02 & $c$ & ATLAS \\
				57844.56 & 11.32 & 0.01 & $I$ & OGLE \\
				57844.68 & 14.24 & 0.03 & $V$ & ASAS-SN \\
				57844.68 & 14.21 & 0.03 & $V$ & ASAS-SN \\
				57844.68 & 14.22 & 0.03 & $V$ & ASAS-SN \\
				\dots & \dots & \dots & \dots & \dots\\
				\hline                  
			\end{tabular}
			\tablefoot{This table is available in electronic format in the full form.}
		\end{table}
		
	\end{appendix}
\end{document}